\documentclass[twocolumn,showpacs,preprintnumbers,amsmath,amssymb]{revtex4}

\usepackage{graphicx}
\usepackage{dcolumn}
\usepackage{bm}

\begin{document}


\title{A possible scenario of metallization in boron doped diamond CB$_x$}

\author{Yu.G. Pogorelov}
\altaffiliation{Electronic address: ypogorel@fc.up.pt}
\affiliation{CFP and Departamento de Fisica, Faculdade de
Ciencias, Universidade do Porto, Rua do Campo Alegre, 687,
4169-007 Porto, Portugal}%

\author{V.M. Loktev}
\altaffiliation{Electronic address: vloktev@bitp.kiev.ua}
\affiliation{Bogolyubov Institute for Theoretical 
Physics, Metrologichna str. 14-b, Kiev, 03143 Ukraine}%

\date{\today}

\begin{abstract}
Possibility for collectivization of acceptor states in a semiconductor,
converting it to metal, is discussed within the scope of Anderson s-d 
hybride model. This model is generalized for multicomponent band structure 
and composite acceptor states, localized on pairs of neighbor dopants 
(impurity ``dumbbells''), in order to describe the boron doped diamond CB$_x$. 
The resulting parameters of band structure, in particular, position of the 
Fermi level, are compared to the recent experimental data on metallized and 
superconducting CB$_x$.
\end{abstract}



\maketitle

\noindent Recent discovery of superconducting transition in boron doped 
diamond CB$_x$ \cite{ekimov} brought a new attention to the problem of 
impurity induced metallization in semiconductors. This topic is already at 
the center of discussion on high-$T_{c}$ metal-oxide perovskites (HTSC), 
where shallow acceptors by, e.g., divalent alcali earth substitutes A for 
trivalent lanthanum in La$_{2-x}$A$_x$CuO$_{4}$, give rise to metallic and 
superconducting state in basal CuO$_{2}$ planes at doping $x$ above some 
critical level $\sim 5\%$. A description of this process on the basis of 
Lifshitz model of impurity states was proposed by the authors \cite{lp}. 
However the situation in CB$_x$ differs from HTSC in essential aspects: 
diamond has an exemplary 3D isotropic (cubic) lattice structure \cite{herman} 
and B atoms at relevant doping levels ($x\sim 4\%$) mostly occupy interstitial 
positions in it \cite{gild},\cite{prins},\cite{chen},\cite{thonke} where they 
nominally should stay neutral. Even at lower doping ($x\lesssim 0.5\%$), 
when they are mostly acceptor substitutes \cite{werner}, it was already 
recognized that common effective mass approach does not apply for the deep 
0.37 eV level by isolated acceptor \cite{glez} whereas the conductivity 
may result from the interplay between that and some ``additional'', much 
more shallow levels ($\sim 0.06$ eV) by acceptor clusters \cite{mamin}. 

\begin{figure}
\includegraphics[scale=0.5]{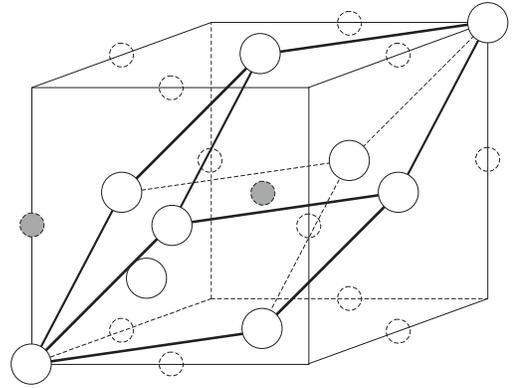}
\caption{\label{cap:1} Unit cell of diamond (bigger circles for carbons) 
where boron dopants (dark circles) occupy an interstitial site at the center 
and one of 12 nearest neighbor interstices (dashed circles). The resulting 
impurity ``dumbbell'' has two acceptor levels, one of them, shallow 
antibonding, being of principal importance for metallization.}
\end{figure}

This clustering effect should be even more important at higher doping, as seen 
from statistical weights of configurations around an interstitial impurity 
(Fig.\ref{cap:1}) at $x = 4\%$ ($8\%$ per diamond unit cell), reaching maximum 
$\sim 38\%$ for its clusters with another impurity in one of 12 nearest 
neighbor interstitices. The facts that carrier density still grows with 
doping and that energy would be gained when doped holes are spread around the 
pair of impurities by B-C covalency effects suggest that such clustered B 
interstitials (impurity ``dumbbells'') also give rise to shallow acceptor 
levels, supposedly not described by effective mass, like those from clusters 
of substitutional impurities. 

When choosing an adequate model for such perturbation of electronic spectrum, 
one have to rule out either the Mott metallization (since the effective 
Bohr radius is too small, $\sim 3$ \AA \cite{baskaran}) and the Lifshitz 
impurity model (which does not provide metallization in 3D doped systems 
\cite{iv}). Then the most natural choice is perhaps the  Anderson s-d hybride 
model \cite{and}, where the possibility for metallization was studied long 
ago \cite{ivanov}. Here we try to adapt the latter results for the intriguing 
physics of doped diamond, leaving aside the issue of pairing mechanism in 
the metallized CB$_x$ (discussed recently in RVB \cite{baskaran} or 
bond-stretching \cite{lee} scenarios).

Referring to the well established theoretical band structure of pure diamond 
\cite{herman},\cite{cohen} and recent ARPES measurement data \cite{jimenez},
the analysis should concentrate around the top of the valence band (chosen
as the energy reference). Here the three valence subbands have almost 
isotropic dispersion $\varepsilon _{j,{\bf k}}=-\hbar^2 k^2/2m_j$ with 
effective masses $m_{1,2,3}\approx 2.12, 1.06, 0.7 m_e$ (neglecting a small 
spin-orbit splitting down by $\approx 6$ meV for $j=2$). Considering the most
relevant acceptor level $\varepsilon_i$ by impurity cluster, closely atop 
the valence band, and neglecting all other (much higher or/and much fewer) 
levels by clustered or isolated impurities, the corresponding generalization 
of hybride model can be presented by the Hamiltonian 
\begin{eqnarray}
H=\sum_{j,{\bf k}}&&\varepsilon _{j,{\bf k}}a_{j,{\bf k}}^{\dagger }
a_{j,{\bf k}}+\sum_{{\bf p}}\varepsilon _{0}b_{{\bf p}}^{\dagger }
b_{{\bf p}}+\notag\\&&+\frac 1{\sqrt{N}}\sum_{j,{\bf k,p}}\left( \gamma_j 
{\rm e}^{i{\bf k}\cdot{\bf p}}a_{j,{\bf k}}^{\dagger }b_{{\bf p}}+
h.c.\right) ,  \label{eq1}
\end{eqnarray}
where $N$ is the number of unit cells in crystal, $a_{j,{\bf k}}$ and 
$b_{{\bf p}}$ are the Fermi operators (spin indices suppressed) for 
excitations of $j$th subband and of impurity clusters (called simply 
impurities in what follows) randomly located at ${\bf p}$ with concentration 
$c \approx x/2$ (per unit cell), and $\gamma_j$ is the constant of 
hybridization between them \cite{note}. 

The resulting spectrum generally includes either band-like and localized 
states and, since each of $cN$\ impurities provides one (hole) carrier to 
the system, the issue of metallization depends on whether the Fermi level 
$\varepsilon _{{\rm F}}$ from the equation  
\begin{equation}
c=\int_{\varepsilon _{{\rm F}}}^{\infty }\rho \left( \varepsilon \right)
d\varepsilon   
\label{eq2}
\end{equation}
lies within the domain of band states. The density of states (DOS, per unit 
cell) $\rho\left(\varepsilon\right) =N^{-1}{\rm{Im Tr}}\left(\varepsilon -
H\right)^{-1}$  in the relevant basis of states for Eq.\ref{eq1}: 
$\left\{\left|j,{\bf k}\right\rangle ,\left|{\bf p}\right\rangle\right\}$ 
(ignoring the conduction band), is calculated through the respective diagonal 
Green functions. 

Let's begin supposing that only single (say, 1st) subband is hybridized with 
impurity states: $\gamma_{2,3}=0$, $\gamma_1=\gamma\neq 0$ (comparison with an 
alternative choice $\gamma_1=\gamma_2=\gamma_3=\gamma$ will be done at the end),
and obtain DOS as a sum of contributions from host (h) and impurity (i) 
subsystems: 
\begin{eqnarray*}
\rho\left(\varepsilon\right)&=&\rho_{h}\left(\varepsilon\right)
+\rho_{i}\left(\varepsilon\right) , \\
\rho _{h}\left( \varepsilon \right)&=&\sqrt{-\varepsilon}\left(W_2^{-3/2} 
+ W_3^{-3/2}\right) + \\&&+\frac 1{\pi N}{\rm {Im}}\sum_{\bf k}\frac 1 {\varepsilon -\varepsilon _{1,{\bf k}}- \Sigma_{{\bf k}}}, \\
\rho_{i}\left( \varepsilon \right)  &=&\frac c{\pi}{\rm {Im}}
\frac 1{\varepsilon -\varepsilon _{0}-\Sigma _{i}}.
\end{eqnarray*}
Here $W_j = \pi^{4/3}\hbar^2/(m_j v^{2/3})$ ($v$ being the unit cell volume) 
is of the order of width of respective subband, whereas the self-energies 
$\Sigma_{\bf k}$ and $\Sigma_i$ can be presented as expansions in groups of 
interacting impurities \cite{ivanov}:
\begin{equation*} 
\Sigma _{{\bf k}}=\frac{c\gamma ^{2}}{\varepsilon-\varepsilon _{0}-
\Sigma _{i}}\left( 1+cB_{{\bf k}}+\ldots\right) 
\end{equation*}
and 
\begin{equation*}
\Sigma _{i}=\frac{\gamma ^{2}}{N}\sum_{{\bf k}}\frac 1 {\varepsilon
-\varepsilon _{1,{\bf k}}-\Sigma _{{\bf k}}}\left(
1+cB_{i}+\ldots \right).
\end{equation*}
The relevant impurity level is defined by the equation $\varepsilon_i = 
\varepsilon_0 + {\rm Re} \Sigma_i(\varepsilon_i)$ in the limit $c \to 0$, 
whereas the next terms after unity in the group expansions:
\begin{eqnarray*}
B_{{\bf k}}&=&-A_{\bf 00}-A_{\bf 00}^{2}+\sum_{{\bf n}\neq {\bf 0}}
\frac{A_{\bf 0n}^3 {\rm e}^{-i{\bf k} \cdot {\bf n}}+A_{\bf 0n}^4}
{1-A_{\bf 0n}^2}, \\
B_{i}&=&-A_{\bf 00}^{2}+\sum_{{\bf n}\neq {\bf 0}}\frac{A_{\bf 0n}^{4}}{1-
A_{\bf 0n}^{2}}, \\
A_{\bf 0n}&=&\frac{\gamma^2}{N\left(\varepsilon -\varepsilon _{0}-
\Sigma _{i}\right)}\sum_{{\bf k}}\frac{{\rm e}^{i{\bf k}\cdot{\bf n}}}
{\varepsilon -\varepsilon _{1,{\bf k}}-\Sigma _{{\bf k}}}
\end{eqnarray*}
describe all indirect interactions between two impurities (in fact, 
two ``dumbbels'') located at cells ${\bf 0}$ and ${\bf n}$, via exchange 
of virtual excitations from 1st subband. The respective contribution to 
DOS from the states localized at such pairs of ``dumbbells'' follows from 
the formula: ${\rm Im} \sum_{\bf n} f_{\bf n}/(1 - A_{\bf 0n}^{2}) \approx 
(\pi/v)\int d{\bf r} f_{\bf r}\delta(1 - {\rm Re}A_{\bf 0r}^{2})$ 
(for any real function $f_{\bf n}$), and the concentration broadening 
$\Gamma_i$ of the impurity level $\varepsilon_i$ from the condition: 
$|{\rm Re}A_{{\bf 0}\overline{\bf r}}(\varepsilon_i \pm \Gamma_i)| = 1$, 
where $\overline{r}\sim (v/c)^{1/3}$ is the average distance between 
impurities. 

\begin{figure}
\includegraphics[scale=0.5]{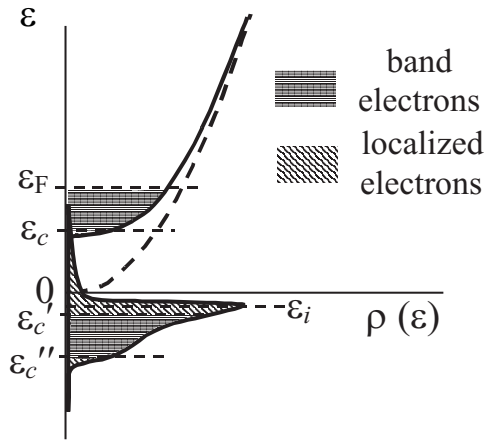}
\caption{\label{cap:2} Restructuring of the conduction band at high 
enough doping ($c \gg c_{cr}$, see in text) by shallow donors within 
the hybride model \cite{ivanov}. There are two separate regions of band 
states near the bottom of conduction band and the system turns metal.}
\end{figure}

We construct the solution by analogy with the case of donor impurity 
level $\varepsilon _i < 0$ hybridized to a single parabolic band 
$\varepsilon _{{\bf k}}=\hbar ^{2}k^{2}/2m$ \cite{ivanov}. There a 
qualitative restructuring of spectrum happens when the concentration 
$c$ surpasses the characteristic (supposedly small) value 
$c_{0}\sim \left( \varepsilon_{i}/W\right)^{3/2}/4\pi$, where 
$W = \pi^{4/3}\hbar^2/(m v^{2/3})$. In other words, then $\overline{r}$ 
turns smaller of the localization radius $r_{0}\sim v^{1/3}\sqrt{W/
\varepsilon _i}$ and localized impurity states effectively overlap. 
If $c$ is also greater than other characteristic value $c_{cr}\sim 
\left(\gamma /W\right)^6$, this overlapping leads to formation of two 
separate bands of coherent extended states, whose structure depends on 
a specific relation between $c_0$ and $c_{cr}$. In particular, if the level 
$\varepsilon_i$ is so shallow that $c_0 < c_{cr}$, the dispersion 
of two split bands is
\begin{equation}
\varepsilon _{\pm}\left({\bf k}\right) \approx \frac{\varepsilon _{\bf k}\pm
\sqrt{\varepsilon _{\bf k}^2+4c\gamma^2}}2,
\label{eq3}
\end{equation}
for wave numbers $k$ restricted to $k \gtrsim k_{min}\sim (c_{cr}/v)^{1/3}$ 
for the upper band, $\varepsilon _{+}\left({\bf k}\right)$, and to 
$k_{min}\lesssim  k \lesssim k_{max} \sim 1/\overline{r}$ for the lower band, 
$\varepsilon _{-} \left({\bf k}\right)$. These restrictions give estimates for 
the Mott mobility edges separating band and localized states: $\varepsilon_c 
\sim \varepsilon_{+}(k_{min})$, $\varepsilon_c^\prime \sim \varepsilon_-(k_{max})$, 
$\varepsilon_c^{\prime\prime}\sim\varepsilon_-({k_{min}})$ (Fig.\ref{cap:2}). Then 
it can be shown that the lower band only adopts $\sim Nc^{3/4}/c_{cr}^{1/4}$ 
states, that is less than the total of $Nc$ carriers. Hence the rest of 
carriers should occupy the ''tail'' states formed by interacting impurity pairs 
whose density is defined as
\begin{equation}
\rho _{i}\left(\varepsilon \right) \sim \frac {c^2 \gamma^2}{\varepsilon 
^2}{\rm Im} B_{i}\sim \frac {c^2 \gamma^8}{W^3\varepsilon^5}
\label{eq4}
\end{equation}
(for $\varepsilon - \varepsilon_i \gg \Gamma_i \sim c^{1/3}\gamma^2/W$) and 
extends to the edge of the upper band. At $\varepsilon > \varepsilon_+(0)$, 
the DOS in this band $\rho_h(\varepsilon)$  is much higher than the ``tail'' 
function $\rho _{i}\left(\varepsilon \right)$, Eq.\ref{eq4}, so that the number 
of empty tail states above the Fermi level approximately equals the number of 
hole carriers within the upper band:
\begin{equation} 
\int_{\varepsilon _{{\rm F}}}^{\infty }\rho
_{i}\left( \varepsilon \right) d\varepsilon \approx \int_{\varepsilon_+(0)}
^{\varepsilon _{\rm F}}\rho_{h}\left(\varepsilon\right) d\varepsilon.
\end{equation}
This gives the distance from the upper band edge to the Fermi energy: 
$\varepsilon _{{\rm F}}- \varepsilon_+(0) \sim \sqrt{c}\gamma$, which is 
$\sqrt{c/c_{cr}}$ times bigger than the distance from this band edge to 
the mobility edge $\sim c_{cr}^{2/3}W$. Therefore the system turns to be 
metal with the band structure as shown in Fig.\ref{cap:2}. 

\begin{figure}
\includegraphics[scale=0.5]{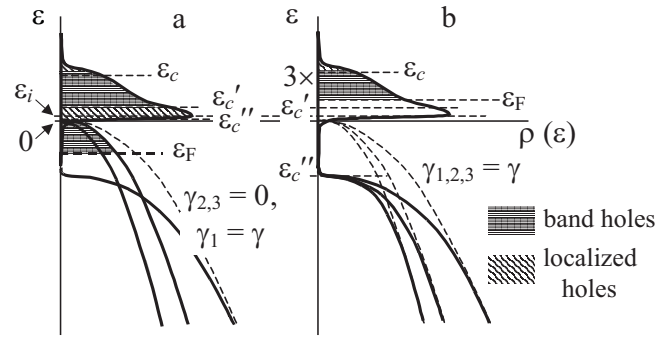}
\caption{\label{cap:3} Restructuring of the multicomponent valence band 
at doping by shallow acceptors. a) Exclusive hybridization to the 1st subband. 
b) Equal hybridization to all three subbands.}
\end{figure}

In the multicomponent valence band case defined by Eq.\ref{eq1} with 
$\gamma_1=\gamma$ and $\gamma_{2,3}\approx 0$, we obtain the same splitting 
as by Eq.\ref{eq3} for the $j=1$ valence subband (within to obvious inversion 
$\varepsilon_+\leftrightarrows\varepsilon_-$), but the essential difference 
from the situation in Fig.\ref{cap:2} is that here the excess carriers can 
fill the two subbands, \emph{non-hybridized to impurities}, resulting in the 
metallic state with the Fermi level below zero energy: $-\varepsilon_{\rm F}
\sim (c-c^{3/4}c_{cr}^{1/4})^{2/3}W$, and two close regions of high occupation, 
as shown in Fig.\ref{cap:3}a. These two regions can play the same role for 
the non-monotonic temperature dependence of conductivity as two activation 
energies, considered in the low doping regime \cite{mamin}. With tentative 
values $\gamma \approx 1$ eV, $\varepsilon_i \approx 0.05$ eV, and 
$W_1\approx 3.5$ eV, we obtain from the above formulas: $c_{0}\sim 0.02\%$, 
$c_{cr}\sim 0.15\%$, and for $c\approx 2\%$ arrive at $|\varepsilon _{{\rm F}}| 
\sim 150$ meV. This value turns comparable to the splitting $\sqrt{c}\gamma$ 
of the 1st subband and quite sufficient to accomodate possible superconducting 
gap $2\Delta \sim 3.52T_{c}\sim 0.6$ meV (for the reported $T_{c}=2.3$ K in 
CB$_x$ \cite{ekimov}). 

An alternative hybridization scheme, with $\gamma_{1,2,3}=\gamma$, will 
produce similar splitting of all three valence subbands, so that the total 
capacity of upper split bands (about triple of that for the single 
$\varepsilon_-({\bf k})$ band in Eq.\ref{eq3}) will be most probably 
sufficient to accomodate the Fermi level at positive energies, as shown 
in Fig.\ref{cap:3}b. The value of $\varepsilon_+(0)-\varepsilon _{{\rm F}}$ 
in the resulting conduction band should be comparable to the above estimate for 
$|\varepsilon _{{\rm F}}|$, but this single region of high occupation looks 
unlike for the above referred non-monotonic conductivity vs temperature, 
seen also in the metallized CB$_x$ \cite{ekimov}. This favors to the former 
version of impurity preferential hybridization to a single valence subband. 
 
In conclusion, we analyzed the specifics of impurity induced metallization
in a version of Anderson s-d hybride model for multiband semiconductor and
showed its difference compared to the simple single band case. The resulting
split band structure with a relatively small Fermi surface of radius 
$k_{{\rm F}}\sim \left(c/v\right) ^{1/3}$ near the Brillouin zone
center also essentially differs from the non-split structures obtained in the
simplest appoximations of effective mass or virtual crystal \cite{lee}. This 
difference can be verified experimentally, e.g., in the optical response at 
about 8 $\mu$m. At least, the proposed metallic state in CB$_x$ can be further 
used to describe the superconducting transition in such doped system, which, 
regardless of the pairing mechanism (perhaps phonon-mediated), can be 
effectively enhanced due to multiband spectrum structure \cite{mosk}.


\begin{thebibliography}{99}
\bibitem{note} It should be noted that, strictly speaking, there are 3 
different types of the considered impurity dumbbels, related to their 
orientations along $x$, $y$, or $z$ crystalline axes. However, we do not 
expect any specific asymmetry between them in the cubic crystal and attribute 
the same $\varepsilon_0$ and $\gamma_j$ values to all of them.
\end{thebibliography}
\end{document}